\documentclass[twocolumn,english,english,english,aps]{revtex4}
\usepackage[T1]{fontenc}
\usepackage[latin1]{inputenc}

\makeatletter

\providecommand{\LyX}{L\kern-.1667em\lower.25em\hbox{Y}\kern-.125emX\@}

\usepackage[T1]{fontenc}
\usepackage[latin1]{inputenc}
\usepackage{babel}

\makeatletter

\usepackage[T1]{fontenc}
\usepackage[latin1]{inputenc}
\usepackage{babel}



\usepackage{babel}
\makeatother
\begin{document}

\title{Phantom energy mediates a long-range repulsive force}

\author{Luca Amendola}

\affiliation{INAF/Osservatorio Astronomico di Roma, \\
Viale Frascati 33, 00040 Monteporzio Catone (Roma), Italy }

\begin{abstract}
Scalar field models with non-standard kinetic terms have been proposed
in the context of $k$-inflation, of Born-Infeld lagrangians, of phantom
energy and, more in general, of low-energy string theory. In general,
scalar fields are expected to couple to matter inducing a new interaction.
In this paper I derive the cosmological perturbation equations and
the Yukawa correction to gravity for such general models. I find three
interesting results: first, when the field behaves as phantom energy
(equation of state less than $-1$) then the coupling strength is
negative, inducing a long-range repulsive force; second, the dark
energy field might cluster on astrophysical scales; third, applying
the formalism to a Brans-Dicke theory with general kinetic term it
is shown that its Newtonian effects depend on a single parameter that
generalizes the Brans-Dicke constant.
\end{abstract}

\date{{\today}}

\maketitle
There has been in recent years a renewal of interest in scenarios
that propose alternatives or corrections to Einstein's gravity. These
proposals are rather heterogeneous: they find their roots in multidimensional
theories in which gravity propagates in more than three dimensions
\cite{ah}, in scalar fields predicted in superstrings that mediate
long-range interactions \cite{anto}, in various mechanisms that might
give cosmologically observable effects \cite{wet88} and violate the
equivalence principle \cite{dam} or Newton's inverse square law (see
e.g. \cite{sundrum}). Moreover, the cosmological observations of
dark energy \cite{riess} can be explained in terms of an almost massless
scalar field that, if coupled to matter, would again modify gravity
\cite{wet88,dam,wet95,ame3,bacc00}. Although the scale of the dark-energy
interaction is expected to be of cosmological size, observable effects
in clusters \cite{maccio} and galaxies \cite{mari} have been investigated.

Along with the increase of interest in the theory, new experimental
tests of gravity have been performed (see the reviews in refs. \cite{fish}
and recent results in \cite{hoyle}). Most laboratory experiments
interpret the results in terms of a possible Yukawa addition to the
gravitational potential \begin{equation}
V(r)=-G\frac{m_{1}m_{2}}{r}\left[1+\alpha e^{-r/\lambda _{s}}\right]\, .\label{eq:yuk}\end{equation}
This form is the static limit of interactions from exchange of bosons
of mass $m_{s}=\lambda _{s}^{-1}$ (in Planck units); the interaction
strength $\alpha $ is proportional to the square of the couplings.
In view of the results below, it is important to remark that for spin-$0$
particles with a standard Lagrangian the strength $\alpha $ is always
positive. A negative amplitude $\alpha $ (i.e. a repulsive effect)
can be obtained only via the exchange of vector bosons \cite{fish}
or in models with finite size of gravitons \cite{sundrum}. The sign
of $\alpha $ gives therefore an important indication of the underlying
physics.

Aim of this paper is to evaluate the Yukawa correction to gravity
when the scalar field $\phi $ has a non-standard kinetic term, i.e.
when the action is generalized as follows ($c=8\pi G=1$, signature
-+++):\begin{equation}
S=\int d^{4}x\sqrt{-g}[\frac{R}{2}+p(X,U(\phi ))]+S_{m}(\rho _{m},e^{2\omega (\phi )}g_{\mu \nu })\label{eq:lagr1}\end{equation}
where $X=-\phi _{,\mu }\phi ^{,\mu }/2$, $U(\phi )$ is any function
of $\phi $, and where field and matter are coupled in the action
$S_{m}$ through the scalar-tensor metric $e^{2\omega (\phi )}g_{\mu \nu }$;
we are working therefore in the Einstein frame (the subscript $m$
denotes quantities that refer to matter). At the same time, and with
the same formalism, we will derive the equations of cosmological perturbations.
In the standard case, $p=X-U(\phi )$. Then the functions $\alpha $
and $\lambda _{s}$ in the Yukawa correction are \cite{dam-nor}\begin{eqnarray}
\alpha  & = & 2(d\omega (\phi )/d\phi )^{2}\equiv 2\omega _{,\phi }^{2}\, ,\nonumber \\
\lambda _{s} & = & m_{s}^{-1}=(d^{2}U/d\phi ^{2})^{-1/2}\equiv (U_{,\phi \phi })^{-1/2}\label{eq:yukstan}
\end{eqnarray}
(there is an extra factor of 2 in $\alpha $ respect to \cite{dam-nor}
because they assume $2X$ as kinetic term).

Higher-order kinetic terms appear quite generically in superstrings
(see e.g. \cite{gpv,piazza,tsu}). Moreover, they have been used in
cosmology to model an accelerated phase without potential, the so
called $k$-inflation or $k$-essence \cite{arme,garriga}. General
kinetic terms can also reproduce the behavior of {}``phantom energy'',
i.e. a field with an equation of state $w=p/\rho <-1$, which is realized
with a negative kinetic term \cite{phantom}. Finally, they have been
introduced also in the form of the Born-Infeld lagrangian $L_{BI}=-U(\phi )(1-2M^{-4}X)^{1/2}$
\cite{fin}. Writing the energy-momentum tensor as\begin{equation}
T_{\mu \nu }^{(\phi )}=g_{\mu \nu }p+p'\phi _{,\mu }\phi _{,\nu }\, ,\label{eq:tmunu}\end{equation}
where $p'=\partial p/\partial X$, one can readily identify the pressure
with $p$ and the energy density with \cite{garriga}\begin{equation}
\rho =2Xp'-p\, .\label{eq:def}\end{equation}
 We find convenient to write the potential as $U=U_{0}e^{-\sqrt{2/3}f(\phi )\phi }$,
so that $U_{,\phi }=-\sqrt{2/3}f_{1}(\phi )U$ and the $n-$th derivative\begin{equation}
U_{,\phi }^{(n)}=(-1)^{n}\left(2/3\right)^{n/2}f_{n}U\: ,\label{eq:}\end{equation}
where $f_{0}=1$ and $f_{n}=f_{n-1}f_{1}-(df_{n-1}/d\phi )/(\sqrt{\frac{2}{3}})\: $
. All general expressions below (i.e. those in which we do not assume
a specific form of $p(X,U)$) could be simplified by a redefinition
of the field $\phi ^{*}\equiv f(\phi )\phi $ such that the potential
becomes exponential; in practice this amounts to setting $f=\mu =const$
and $f_{n}=\mu ^{n}$. The Klein-Gordon equation of motion in a flat
FRW metric with scale factor $a$ written in the $e$-folding time
$\eta =\log a$ is\begin{eqnarray}
x'+\left(3+\frac{H'}{H}\right)x-18Ax^{3}H^{2}p^{2,0}+ &  & \nonumber \\
Ay^{2}f_{1}\mu (p^{0,1}-6x^{2}H^{2}p^{1,1}) & = & A\beta \Omega _{m}\, ,\label{eq:kg}
\end{eqnarray}
where we introduced the variables $x^{2}=\phi '^{2}/6$ and $y^{2}=U/3H^{2}$,
and where $A=(p^{1,0}+6H^{2}x^{2}p^{2,0})^{-1}$ . (Note that $x^{2},y^{2}$
are the kinetic and potential energy density fractions in the standard
case.) We denote with $p^{i,j}$ the derivative $\partial ^{i+j}p(X,U)/\partial ^{i}X\partial ^{j}U$
(but we also use the shorthand $p'=\partial p/\partial X$, $p''=\partial ^{2}p/\partial X^{2}$;
in all other cases the prime denotes $d/d\eta $). The standard case
is recovered when $p^{1,0}=1$, $p^{0,1}=-1$ and $p^{i,j}=0$ for
any $i+j>1$. To conform to previous notation (e.g. \cite{linnonlin}),
we defined instead of $\alpha $ the coupling \[
\beta =(3/2)^{1/2}\omega _{,\phi }\, .\]
The Einstein equations are \begin{eqnarray}
1 & = & \frac{6H^{2}x^{2}p'-p}{3H^{2}}+\Omega _{m}\, ,\nonumber \\
\frac{H'}{H} & = & -\frac{3}{2}\Omega _{m}-3x^{2}p'\, .\label{eq:}
\end{eqnarray}
Notice that $X=3H^{2}x^{2}$. The conservation equation for matter
is the same as in the standard coupled case (e.g. \cite{linnonlin}).
From the definitions of $p$ and $\rho $ we define the field sound
speed \cite{garriga}\begin{equation}
c_{s}^{2}=\frac{\partial p/\partial X}{\partial \rho /\partial X}=\frac{p'}{p'+6H^{2}x^{2}p''}=Ap'\, ,\label{eq:}\end{equation}
and the equation of state \begin{equation}
w=\frac{p}{\rho }=-1+\frac{2x^{2}p'}{1-\Omega _{m}}\, .\label{eq:state}\end{equation}
Phantom dark energy, $w<-1$, requires therefore $p'<0$. For instance,
if $\Omega _{m}\approx 0.3$ and $w\approx -1.3$ \cite{riess} then
$p'\approx -0.1/x^{2}$.

We wish to keep the scalar Lagrangian as general as possible. However,
it is instructive to derive the conditions of classical and quantum
stability. The classical stability is ensured by the condition $\rho >0$
and $c_{s}^{2}>0$; the quantum stability by the positivity of the
Hamiltonian, i.e. \cite{piazza} $A>0$, $p'>0$ and $p^{0,2}<0$.
All together, the conditions are $p<\frac{p'}{2X}$, $p'>0$, $p''>-\frac{p'}{2X}$,
$p^{0,2}<0$.

Generally speaking, the scalar field can couple to matter with different
couplings \cite{dam}. The observational constraints on the baryon
coupling are rather tight\cite{hagi02}, while those on dark matter
are much weaker and depend on the cosmic evolution itself \cite{aq}.
We assume in this paper a simplified model containing only one species
of pressureless matter. 

To derive the weak field limit, we use the longitudinal gauge metric
in $\eta =\log a$:\begin{equation}
ds^{2}=-(1+2\Psi )\frac{d\eta ^{2}}{H^{2}}+e^{2\eta }(1-2\Phi )dx_{i}dx^{i}.\label{eq:metric}\end{equation}
Moreover we redefine $\delta \phi =\sqrt{6}\varphi $ and introduce
the matter density contrast $\delta _{m}=\delta \rho _{m}/\rho _{m}$
, the dimensionless matter velocity divergence $\theta _{m}=\nabla _{i}v_{(m)i}/H$
and the field density contrast\begin{equation}
\Omega _{\phi }\delta _{\phi }\equiv \frac{\delta \rho }{3H^{2}}=2\varphi y^{2}f_{1}\mu (p^{0,1}-6x^{2}H^{2}p^{1,1})-\frac{2x^{2}}{A}\Phi +\frac{2x}{A}\varphi '\, .\label{eq:}\end{equation}
In absence of anisotropic stress $\Phi =\Psi $ and the metric perturbation
in Fourier space with wavelength $\lambda =Ha/k$ (not to be confused
with $\lambda _{s}$) is \begin{equation}
\Phi =-\frac{3}{2}\lambda ^{2}[(\delta _{m}+3\lambda ^{2}\theta _{m})\Omega _{m}+\delta _{\phi }\Omega _{\phi }+6x\varphi p^{1,0}]\, .\label{eq:phinp}\end{equation}
From now on we assume $\beta =const$ although the generalization
to $\beta (\phi )$ is straightforward (see for instance \cite{linnonlin}).
The matter equations are\begin{eqnarray}
\delta _{m}' & = & -\theta _{m}+3\Phi '-2\beta \varphi '\: ,\\
\theta _{m}' & = & -\left(2+\frac{H'}{H}-2\beta x\right)\theta _{m}+\lambda ^{-2}(\Phi -2\beta \varphi )\: .\label{cdm-n}
\end{eqnarray}

Let us now go to the Newtonian limit of small scales, i.e. $k\to \infty $
or $\lambda \to 0$. At the first non-trivial order the perturbed
scalar field equation is\begin{equation}
\varphi ''+F(\phi )\varphi '+\hat{m}(\phi )^{2}\varphi +c_{s}(\phi )^{2}\lambda ^{-2}\varphi =A\beta \Omega _{m}\delta _{m}\, .\label{eq:pertkg}\end{equation}
 The function $F$ reduces to $(3+H'/H)$ in the standard case and
in the static limit $x\to 0$; its general expression depends on higher
order derivatives, up to $p^{3,0}$ , but is very complicated and
unuseful for what follows (detailed calculations will be published
elsewhere). The dimensionless effective mass $\hat{m}$ in units of
$H^{-1}$ is given by\begin{eqnarray}
A^{-1}\hat{m}^{2} & = & 2y^{2}p^{0,1}(3ABf_{1}^{2}H^{2}y^{2}-f_{2})\nonumber \\
 & + & 3x^{2}p^{1,0}(3p^{1,0}+A^{-1})\nonumber \\
 & - & 6f_{1}^{2}H^{2}y^{4}p^{0,2}+18ABH^{2}xy^{2}f_{1}p^{1,0}\nonumber \\
 & - & 6H^{2}y^{2}[6ABf_{1}^{2}H^{2}x^{2}y^{2}p^{1,1}+ABf_{1}\beta \Omega _{m}\nonumber \\
 & + & xp^{1,1}(3f_{1}-2f_{2}x)-6f_{1}^{2}H^{2}x^{2}y^{2}p^{1,2}]\, .\label{eq:mass}
\end{eqnarray}
 where $B=p^{1,1}+6H^{2}x^{2}p^{2,1}$. Consider now eq. (\ref{eq:pertkg})
in absence of matter, $\Omega _{m}=0$, i.e. for a self-gravitating
boson fluid. For a standard kinetic term, $c_{s}=1$ and $\hat{m}^{2}=U_{,\phi \phi }/H^{2}-12x^{2}$.
The effective mass is of order unity or less if the field is identified
as dark energy (since $x^{2}\le 1$ from the constraints on $w$,
see eq. (\ref{eq:state}), and $U$ is expected to be slow-rolling
with respect to the expansion rate): therefore the growth of the scalar
field fluctuations is prevented at all sub-horizon scales. Scalar
field clustering is of course possible for large and negative potentials
or rapid oscillations \cite{matos}, adding internal degrees of freedom
\cite{lesg}, and in general at near-horizon scales \cite{wet2002}.
In the general kinetic case, however, different values and signs of
$c_{s}$ and $\hat{m}$ are possible. If $\hat{m}^{2}$ is negative,
then the field fluctuations grow at scales (in units of $H^{-1}$)
$\lambda >\hat{\lambda }_{s}=|c_{s}/\hat{m}|$; if $c_{s}^{2}<0$,
the instability is at scales smaller than $\hat{\lambda }_{s}$; finally,
if both are negative, then the instability is at all sub-horizon scales. 

Let us restrict now the attention to the case $\hat{m}^{2},c_{s}^{2}>0$.
Then the solution of the homogeneous equation associated to (\ref{eq:pertkg})
is a fast oscillation around zero; the amplitude of $\varphi $ is
driven by the rhs forcing and is given by the solution of\begin{equation}
\hat{m}(\phi )^{2}\varphi +c_{s}(\phi )^{2}\lambda ^{-2}\varphi =A\beta \Omega _{m}\delta _{m}\, ,\label{eq:}\end{equation}
that is \begin{equation}
\varphi =\frac{A\beta \Omega _{m}\delta _{m}a^{2}H^{2}}{\hat{m}^{2}a^{2}H^{2}+c_{s}^{2}k^{2}}=\frac{\beta }{p'}\frac{k^{2}\lambda ^{2}\Omega _{m}\delta _{m}}{a^{2}\left(\frac{\hat{m}^{2}H^{2}}{c_{s}^{2}}\right)+k^{2}}\, .\label{eq:phitot}\end{equation}
 From the equation for $\theta _{m}'$ and from (\ref{eq:phitot})
we see that the potential acting on matter is the Newton-Yukawa potential
$\Phi ^{*}=\Phi -2\beta \varphi $ that, upon Fourier transform, obeys
the Poisson equation\begin{equation}
\nabla _{i}\nabla ^{i}\Phi ^{*}=-\frac{3}{2}H^{2}\Omega _{m}\delta _{m}(1+\tilde{\alpha }e^{-r/\tilde{\lambda }})\: ,\label{eq:newpot}\end{equation}
where \begin{eqnarray}
\tilde{\alpha } & = & \frac{4}{3}\frac{\beta ^{2}}{p'}=\frac{\alpha }{p'}\, ,\label{eq:atilde}\\
\tilde{\lambda } & = & \frac{\hat{\lambda }_{s}}{H}=\frac{c_{s}}{\hat{m}H}\, .\label{eq:mtilde}
\end{eqnarray}
These expressions generalize eqs. (\ref{eq:yukstan}) and are the
main result of this paper. From $\tilde{\alpha }$ we see that a phantom
field, $p'<0$, coupled to matter mediates a \emph{repulsive} force,
contrary to bosons with standard kinetic terms whose interaction is
always attractive. The force is long range if $\hat{m}\le c_{s}$:
this depends on the particular lagrangian but for a field that drives
the cosmic acceleration this is what one expects. In general, the
repulsive scalar gravity may balance and overcome gravity at small
scales. From the general expressions for $\tilde{\alpha }$ and $\tilde{\lambda }$
we derive now two limits that we can call the \emph{laboratory} limit
(i.e. the static interaction) \emph{}and the \emph{cosmological} limit
(i.e. a massless or {}``dark energy'' field ). Finally, we apply
the results to the Brans-Dicke model.

\emph{Laboratory limit.} We specify the Yukawa correction to the case
in which the background solution is static, $\phi ''=\phi '=0$, i.e.
$x\to 0$ (assuming that $p$ does not diverge, e.g. that it does
not contain inverse powers of $X$). Then we have $c_{s}=1$ and\begin{eqnarray}
\hat{m}^{2} & = & \frac{\hat{m}_{s}^{2}}{p^{1,0}}g(p)\, ,\label{eq:}\\
g & \equiv  & 3\frac{H^{2}f_{1}}{f_{2}}\left[y^{2}f_{1}\left(\frac{p^{1,1}p^{0,1}}{p^{1,0}}-p^{0,2}\right)-\frac{\beta \Omega _{m}p^{1,1}}{p^{1,0}}\right]-p^{0,1}\, ,\nonumber 
\end{eqnarray}
where $\hat{m}_{s}^{2}=2y^{2}f_{2}=U_{,\phi \phi }/H^{2}$ is the
\emph{dimensionless} standard mass, and therefore \begin{equation}
\tilde{\lambda }^{-2}=\frac{U_{,\phi \phi }}{p'}g(p)\, .\label{eq:}\end{equation}
 In general, $g(p)$ can be given any value. For instance, if\begin{equation}
p(X,U)=P_{1}(X)+UP_{2}(X/U)\, ,\label{eq:bdp}\end{equation}
where $P_{1,2}$ are any function for which $P_{1,2}(0)=0$, then
$g=0$ (i.e. an infinite-range force) in the static limit. If instead
the lagrangian $p$ is in the form $p=K(X)-U$ (additive lagrangian),
then $g=1$ and $\tilde{\lambda }=\sqrt{p'}/m_{s}$. It follows that
the effect of the generalized kinetic term on the Yukawa correction
can be neatly absorbed into the two Yukawa parameters $\alpha ,\lambda $
as $\tilde{\alpha }=\alpha /p'$ and $\tilde{\lambda }^{2}=\lambda _{s}^{2}/p'$.
A phantom field with an additive lagrangian and small kinetic energy
$x$ therefore has an imaginary effective mass, unless of course $U_{,\phi \phi }$
is negative as well. As already remarked this implies an instability
of the dark energy fluctuations (whether or not the field is coupled).
It may be conjectured that if such an instability leads to non-linearity
at large scales before the final singularity ({}``big rip'') then
the singularity itself might be avoided. 

\emph{Cosmological limit.} If the field $\phi $ is a dark energy
field its effective mass $\hat{m}$ is expected to be negligible.
Although in the general case this is not necessarily true, we adopt
now this simplification. Then from (\ref{eq:phitot}) we have

\begin{equation}
\varphi =\frac{\beta }{p'}\lambda ^{2}\Omega _{m}\delta _{m}\, .\label{eq:phidelta}\end{equation}
 From this we see that the Yukawa correction to Newton's constant
is $G=G_{N}(1+\frac{4}{3}\frac{\beta ^{2}}{p'})$. Derivating $\delta _{m}'$
and inserting (\ref{eq:phidelta}) we obtain finally

\begin{equation}
\delta _{m}''+\delta _{m}'(2+\frac{H'}{H}-2x\beta )-\frac{3}{2}\Omega _{m}\delta _{m}(1+\frac{4}{3}\frac{\beta ^{2}}{p'})=0\, .\label{eq:matter}\end{equation}
This equation generalizes the similar one given in ref. \cite{linnonlin}
and shows how the kinetic term enters the growth of the matter perturbations.
Although for simplicity it has been derived for $\beta =const$ it
is actually valid also for $\beta =\beta (\phi )$. The solution depends
of course on the cosmological history and on the particular form of
the field Lagrangian. This will be studied in subsequent work. Here
we notice only a few interesting points. First, as already pointed
out, a phantom field has $p'<0$ and therefore its effect on $\delta _{m}$
counteracts gravity. This could be used to set stringent limits on
phantom fields coupled to dark matter. Second, since in general $p'$
evolves in time, then also the term $\beta ^{2}/p'$ is time-dependent,
even if the coupling $\beta $ itself is constant. In particular,
if $p$ contains higher order terms in $X$ then the effective coupling
$\beta ^{2}/p'$ will change when the kinetic terms become sub-dominant,
i.e. when the acceleration sets is. This might introduce novel features
as, e.g., a strong non-gaussianity in the galaxy distribution \cite{ameprl}. 

\emph{Brans-Dicke model}. As an application of the results let us
now consider a generalization of the Brans-Dicke lagrangian\begin{equation}
L=\frac{\sqrt{g}}{2}[\Phi R+\frac{2\omega _{BD}}{\Phi }P(Y)]\, ,\label{eq:bdl}\end{equation}
where $P$ is any function of $Y=-g^{\mu \nu }\Phi _{;\mu }\Phi _{;\nu }/2$
and where $\omega _{BD}$ is the Brans-Dicke constant (a non-constant
$\omega _{BD}$ can be absorbed in $P(Y)$). The Brans-Dicke scalar
field $\Phi $ should not be confused with the gravitational potential
introduced in Eq. (\ref{eq:metric}). Let us call this model kinetic
Brans-Dicke (kBD). Under the conformal transformation $\tilde{g}_{\mu \nu }=e^{2\omega }g_{\mu \nu }\, $
the Lagrangian becomes of the form (\ref{eq:lagr1}) with\begin{equation}
p(X,U)=\frac{X}{3+2\omega _{BD}}\left[3+2\omega _{BD}\frac{U}{X}P(\frac{X}{U})\right]\, ,\label{eq:pbd1}\end{equation}
 if $\Phi =e^{2\omega }$ and if we define $X=-\omega _{;\mu }\omega ^{;\mu }(3+2\omega _{BD})\, ,$$U=(3+2\omega _{BD})e^{-6\omega }/2\, ,$
and $\omega (\phi )=\phi [2(3+2\omega _{BD})]^{-1/2}$. It follows
finally from eq. (\ref{eq:atilde}) that\begin{equation}
\tilde{\alpha }=\frac{2\omega _{,\phi }^{2}}{p^{1,0}}=\frac{1}{3+2\omega _{BD}P_{,Y}}\, ,\label{eq:kbd}\end{equation}
where $P_{,Y}=dP/dY$. Moreover, since $p(X,U)$ has the form (\ref{eq:bdp})
then $\tilde{m}=0$ in the static limit (if $P(Y=0)=0$, for instance
$P=\sum _{n}a_{n}Y^{n}$, $n>0$). This shows that the Newtonian effects
of the kinetic Brans-Dicke theory are parametrized uniquely by $\omega _{kBD}=\omega _{BD}P_{,Y}$
and, therefore, all the experimental constraints on $\omega _{BD}$
in the literature \cite{dam} can be read equivalently as constraints
on $\omega _{kBD}$. A order of unity $\omega _{BD}$ is then acceptable
if $P_{,Y}$ is very large. \emph{}

\emph{Conclusions}. A scalar-tensor metric $e^{2\omega (\phi )}g_{\mu \nu }$
is arguably the simplest, oldest, most studied and most motivated
extension of Einstein's gravity. So far, however, in almost all works
a standard kinetic term has been employed in scalar-tensor theories
(see e.g. \cite{wet88,dam}). Only very recently, refs. \cite{piazza,tsu}
generalized the lagrangian to include both coupling and general kinetic
terms. On the other hand, a generalized kinetic term has been proposed
to account for a phantom equation of state and kinetic-driven inflation
and on theoretical grounds (Born-Infeld lagrangian, higher-order terms
in superstrings). This paper lies at the crossroad of these two streams:
it investigates a kinetic scalar-tensor gravity, focusing on the properties
that do not depend on the particular model. Perhaps the most interesting
result is that a field with equation of state $w<-1$ (a phantom field)
mediates a repulsive force, contrary to ordinary spin-0 bosons. We
also derived the field effective mass in the general case and showed
that it may induce clustering of dark energy. We also have shown that
the Newtonian effects of a kinetic Brans-Dicke model are parametrized
by a single quantity, $\omega _{kBD}=\omega _{BD}P_{,Y}$, that generalizes
the Brans-Dicke constant. 

\emph{Aknowledgments}. I thank M. Gasperini, J. Lesgourgues, F. Piazza,
S. Tsujikawa and G. Veneziano for several useful discussions. I am
indebted to F. Piazza also for discussing ref. \cite{piazza} prior
to publication, which motivated the present work.

\end{document}